\documentclass{aims}

\usepackage[ruled, lined, linesnumbered, commentsnumbered, longend]{algorithm2e}
\usepackage{tabularx, booktabs}
\usepackage{amsmath,amssymb}
\usepackage{bbm}
\usepackage{tabularx,ragged2e}
\usepackage{subcaption,booktabs}
\usepackage{cleveref}
\usepackage{pgfplots}
\pgfplotsset{width=10cm, compat=1.9}
\usepackage{tikz}
\usepackage{biblatex}
\usetikzlibrary{matrix} 
\usetikzlibrary{arrows} 
\usetikzlibrary{calc} 
\usetikzlibrary{external}
\usepackage{txfonts}
\def\typeofarticle{Preprint}

\numberwithin{equation}{section}

\makeatletter
\renewcommand{\@biblabel}[1]{#1\hfill \hspace{-0.2cm}}
\makeatother

\begin{document}
\title{Partitioned Surrogates and Thompson Sampling for Multidisciplinary Bayesian Optimization}

\author{%
  Susanna Baars \affil{1,2}\corrauth
  Jigar Parekh \affil{1,3}
  Ihar Antonau \affil{1,4}
  Philipp Bekemeyer \affil{1,3}
  Ulrich Römer\affil{1,2}
}

\shortauthors{the Author(s)}

\address{%
  \addr{\affilnum{1}}{Cluster of Excellence SE²A – Sustainable and Energy-Efficient Aviation, Technische Universit\"at Braunschweig, Germany}
  \addr{\affilnum{2}}{Institute for Acoustics and Dynamics, Technische Universit\"at Braunschweig, Germany}
  \addr{\affilnum{3}}{Institute of Aerodynamics and Flow Technology, German Aerospace Center, Germany}
  \addr{\affilnum{4}}{Institute of Structural Analysis, Technische Universit\"at Braunschweig, Germany}
}

\corraddr{s.baars@tu-braunschweig.de}

\begin{abstract}

The long runtime associated with simulating multidisciplinary systems challenges the use of Bayesian optimization for multidisciplinary design optimization (MDO). This is particularly the case if the coupled system is modeled in a partitioned manner and feedback loops, known as strong coupling, are present. 

This work introduces a method for Bayesian optimization in MDO called "Multidisciplinary Design Optimization using Thompson Sampling", abbreviated as MDO-TS. Instead of replacing the whole system with a surrogate, we substitute each discipline with such a Gaussian process. Since an entire multidisciplinary analysis is no longer required for enrichment, evaluations can potentially be saved. However, the objective and associated uncertainty are no longer analytically estimated. Since most adaptive sampling strategies assume the availability of these estimates, they cannot be applied without modification.

Thompson sampling does not require this explicit availability. Instead, Thompson sampling balances exploration and exploitation by selecting actions based on optimizing random samples from the objective. We combine Thompson sampling with an approximate sampling strategy that uses random Fourier features. This approach produces continuous functions that can be evaluated iteratively.

We study the application of this infill criterion to both an analytical problem and the shape optimization of a simple fluid-structure interaction example.

\end{abstract}

\keywords{
\textbf{Multidisciplinary Design Optimization $\cdot$ Bayesian optimization $\cdot$ Global optimization $\cdot$ Thompson sampling}
}

\maketitle

\def\sizeInitDoEToyProblem{4}
\def\nIterMDOTSToyProblem{3}

\def\zReferenceToyProblem{-2.9989}
\def\yReferenceToyProblem{9.9607 & 6.9618}
\def\fObjReferenceToyProblem{-1.1495}

\def\zPredInitToyProblem{-5.0}
\def\yPredInitToyProblem{9.525 & 5.3083}
\def\fObjPredInitToyProblem{999.4715}

\def\zPredFinalToyProblem{-2.9997}
\def\yPredFinalToyProblem{9.9509 & 6.9512}
\def\fObjPredFinalToyProblem{-1.1495}

\def\initLengthScale{1}
\def\initScaling{1}
\def\nBasisFunctions{1000}
\def\nugget{$10^{-7}$}
\def\sizeInitDoESellar{5}
\def\nIterMDOTSSellar{10}
\def\nRunsSellar{100}
\def\nConvergedSellar{95}

\def\zReferenceSellar{0. & 2.6345 & 0.}
\def\fObjReferenceSellar{-2.8085}
\def\zPredSellar{0.0001 & 2.6325 & 0.0003}
\def\fObjPredSellar{-2.8072}
\def\zErrorSellar{--- & 0.272 & ---}
\def\fObjErrorSellar{0.0458}

\def\nCSTSide{3}
\def\nCSTTotal{6}
\def\machInfinity{0.2}
\def\initAngleOfAttack{0 rad}
\def\springConstant{$2000$ N rad}
\def\springCoordinates{$\begin{bmatrix} 0, 0 \end{bmatrix}$}
\def\centerOfMass{$\begin{bmatrix} -0.05, 0 \end{bmatrix}$}
\def\CSTRange{$\pm 30 \, \%$}

\def\sizeInitDoEAirfoil{5}
\def\nIterMDOTSAirfoil{10}

\def\CSTInitAirfoil{0.0653 \\ 0.0542 \\ 0.0536 \\ -0.0653 \\ -0.0542 \\ -0.0536}
\def\CSTInitDoE{0.0661 \\ 0.0705 \\ 0.0697 \\ -0.0529 \\ -0.0542 \\ -0.0485}
\def\CSTMDOTS{ 0.0848 \\ 0.0705 \\ 0.0697 \\ -0.0457 \\ -0.038 \\ -0.0375}

\def\angleOfTwistInitAirfoil{-0.0}
\def\angleOfTwistInitDoE{0.0151}
\def\angleOfTwistMDOTS{0.0281}

\def\torqueInitAirfoil{0.0693}
\def\torqueInitDoE{-30.2673}
\def\torqueMDOTS{-56.166}

\def\liftInitAirfoil{-1.3851}
\def\liftInitDoE{605.4142}
\def\liftMDOTS{1123.7555}

\section{Introduction}

Complex multidisciplinary systems are often modeled in a modular manner, where each subsystem is treated as a black-box solver and all subsystems are coupled in a weak or strong way. The optimization of such systems, known as multidisciplinary design optimization (MDO), is challenging. It requires multidisciplinary analysis (MDA), i.e., solving the coupled and often nonlinear system of equations for multiple sets of parameters. Long run times for MDA are common, particularly in strongly coupled systems where feedback loops are present. In order to minimize the associated computational effort, several methods for MDO have been developed and comparatively studied \cite{martins_multidisciplinary_2013}.
In a recent review of large-scale MDO, Hwang et al. concluded that the most efficient approaches to solving MDO problems involve exact gradients computed by the adjoint method \cite{hwang_large-scale_2019}.
However, obtaining the adjoint gradients for complex coupled systems remains challenging. Therefore, current efforts attempt to make gradient-free optimization methods efficient for MDO. In \cite{dubreuil_towards_2020}, Dubreuil et al. propose a Bayesian optimization method where each discipline is replaced by a Gaussian process surrogate instead of the whole system. 

The contributions of this article is the introduction of a novel method that follows this paradigm, referred to as "Multidisciplinary Design Optimization using Thompson Sampling", short MDO-TS. Thereby, we combine coupled surrogates and Thompson sampling based on an approximate sampling scheme known as decoupled sampling. Thompson sampling achieves a trade-off between exploration and exploitation by optimizing random draws from the posterior. The approximate sampling scheme utilizes the random Fourier features (RFF) approximation of Gaussian processes. We apply MDO-TS to an analytical nonlinear unconstrained MDO problem and an engineering test case.

The remainder of this section offers an overview of recent literature along with the essential theoretical background. It is structured as follows: \Cref{subsec:partitioned_surrogates} discusses approaches that replace some or all of the disciplines within a multidisciplinary problem with data-driven surrogates. In \cref{subsec:thompson_bo}, a brief introduction to Gaussian processes and Bayesian optimization is given. Special emphasis is put on Thomson sampling for Bayesian optimization. \Cref{subsec:decoupled_sampling} introduces RFF-based approximate sampling from Gaussian processes.

\subsection{Partitioned data-driven surrogates}\label{subsec:partitioned_surrogates}

Dubreuil et al. propose in \cite{dubreuil_towards_2020} to replace the individual solvers of a multidisciplinary system with Gaussian processes. After initialization, the Gaussian processes are refined using the predictions of the fixed points by the coupled surrogates and a sampling-based expected improvement criterion. The proposed method is called efficient global multidisciplinary design optimization (EGMDO). The performance of EGMDO is evaluated on several analytical examples. The original formulation of EGMDO is only applicable to unconstrained MDO problems, however, \cite{cardoso_constrained_2024} proposes an extension of EGMDO for constrained MDO problems.

Most realistic MDA problems involve high-dimensional coupling variables. The curse of dimensionality hinders the direct applicability of approaches that make use of partitioned surrogates. One approach to solving this problem is to use dimension reduction methods such as proper orthogonal decomposition (POD). This concept has already been proposed in \cite{filomeno_coelho_model_2008}.  Here, moving least squares (MLS) was used to interpolate over the POD coefficients.

In \cite{berthelin_disciplinary_2022}, Berthelin et al. extend some of the concepts of \cite{dubreuil_towards_2020} to numerical models with high-dimensional coupling variables. Therefore, Berthelin et al. propose an iterative scheme for an initial evenly distributed DOE. The model evaluations are used to determine a POD base to reduce the dimensionality of the coupling variables and to build initial Gaussian process surrogates.
\cite{da_costa_cardoso_model_2022} expands on the work of Berthelin et al. by incorporating the sampling-based expected improvement criterion.

To further reduce the number of required model evaluations and to decouple the construction of disciplinary surrogates, \cite{scholten_uncoupled_2021} suggests parameterizing the coupling variables based on a single evaluation of both disciplines instead of applying dimensionality reduction techniques. However, this limits the accuracy of the MDA.

Since only one of the disciplinary models may be expensive to evaluate, it might be beneficial to substitute just that one discipline with a surrogate. In \cite{tiba_non-intrusive_2023}, the authors replace an expensive elastic solid model of a parametrized FSI problem with a surrogate.

While most recent work has focused on Gaussian processes as disciplinary surrogate models for MDA, the application of neural networks has also been explored \cite{gupta_hybrid_2022, arcones_neural_2022}.

\subsection{Gaussian Process Regression for Bayesian Optimization}\label{subsec:thompson_bo}
Surrogate-based optimization is an approach for gradient-free optimization of black-box functions that are costly to evaluate. If a probabilistic surrogate model is used, typically a Gaussian process (GP), this technique is referred to as Bayesian optimization. First, the black-box function is evaluated in an initial design of experiments (DoE). Second, the surrogate is selectively refined according to a strategy dictated by an \textit{acquisition function} or \textit{infill criterion}. Often, the acquisition function strives for a trade-off between exploration and exploitation \cite{frazier_tutorial_2018}. 

Since we will focus on Bayesian optimization with GPs and an infill criterion known as Thompson Sampling (TS), we will now first introduce GPs and then TS. The following introduction to GPs is very brief, as only the nomenclature and symbols used in this article need to be clarified. A detailed introduction to GPs is provided in \cite{rasmussen_gaussian_2005}. 

Consider a domain $D \subset \mathbb{R}^n,$ where $n \geq 1$. GPs are random functions $f:D \to \mathbb{R}$, with a multivariate Gaussian distribution when restricted to a set of points $\{\boldsymbol{x}_1,\ldots,\boldsymbol{x}_n\}$, where $\mathbf{x}_i \in D$. Hence, GPs are fully specified by a mean function $m(\boldsymbol{x})$ and a covariance function or kernel function $k(\boldsymbol{x},\boldsymbol{x}')$ \cite{rasmussen_gaussian_2005}. Data consists of pairs $(\boldsymbol{x}_i,y_i)_{i=1}^n$ and can be incorporated by exact GP inference. More precisely, let $\boldsymbol{K} \in \mathbb{R}^{n \times n}$ denote the covariance matrix with entries $K_{ij} = k(\boldsymbol{x}_i,\boldsymbol{x}_j)$ and let $\boldsymbol{x}_*$ be a location at which we wish to predict the value of $f$. For simplicity, consider a mean-free prior GP, then the posterior mean and covariance are given by 
\begin{align}
    m_{n}(\boldsymbol{x}_*) &= k(\boldsymbol{x}_*,\boldsymbol{X}) \left( \boldsymbol{K} + \sigma_\epsilon^2 \boldsymbol{I} \right)^{-1} \boldsymbol{y}, \\
    k_{n}(\boldsymbol{x}_*,\boldsymbol{x}_*) &= k(\boldsymbol{x}_*,\boldsymbol{x}_*) - k(\boldsymbol{x}_*,\boldsymbol{X}) \left( \boldsymbol{K} + \sigma_\epsilon^2 \boldsymbol{I} \right)^{-1} k(\boldsymbol{X},\boldsymbol{x}_*),
\end{align}
where $\boldsymbol{I}$ denotes the $n \times n$-identity matrix, $\boldsymbol{X}=[\boldsymbol{x}_1,\ldots,\boldsymbol{x}_n]$, $\boldsymbol{y}=(y_1,\ldots,y_n)^\top$ and $k(\boldsymbol{x}_*,\boldsymbol{X}) \in \mathbb{R}^{1 \times n}$ as well as $k(\boldsymbol{X},\boldsymbol{x}_*) \in \mathbb{R}^{n \times 1}$. Moreover, $\sigma_\epsilon^2$ refers to the noise variance, which is assumed to be the same for each data point. The posterior distribution is again a GP, that is fully described by the conditioned mean and covariance function. It should be noted that kernel functions usually contain hyperparameter that are not reflected by the notation, but need to be optimized based on the observed data. The length scale $l$ and the signal variance $\sigma^2$ are common examples for many kernel functions. 

We are interested in settings where the training data comes from an expensive computer model. Hence, it is crucial to acquire the data adaptively based on an effective infill criterion. For Bayesian optimization, the expected improvement (EI) \cite{frazier_tutorial_2018} is a popular choice. However, within this study we will be focusing on TS to explore the possibility of a parallel execution of disciplinary models. TS or probability matching is one of the oldest schemes for balancing exploration and exploitation in multi-armed bandit problems \cite{russo_tutorial_2020, chapelle_empirical_2011}, where actions are determined by optimizing random draws from the posterior distribution. If the outcome of an action is highly uncertain, the random realizations will be highly diverse, encouraging exploration. On the other hand, if an action is known to be highly rewarding with little uncertainty, this will be reflected in all random draws, encouraging exploitation. Applying TS to Bayesian optimization with GPs, here refereed to as GP-TS, is straightforward. At each iteration, we draw a realization of a GP conditioned on the already observed data $(\hat{f}_n|\boldsymbol{y})(\cdot) \sim \mathcal{GP}(m_n, k_n)$. Then we select a new control point according to $ \boldsymbol{x}_j = \arg \min_{\boldsymbol{x}} (\hat{f}_n|\boldsymbol{y})(\boldsymbol{x}) $ \cite{kandasamy_asynchronous_2017}. We will continue to use hat notation for representing random draws throughout this article.

GP-TS sampling offers a flexible probabilistic framework. Hence, it has been utilized to apply Bayesian optimization in challenging scenarios. Examples of such applications can be found in \cite{kandasamy_asynchronous_2017} and \cite{eriksson_scalable_2021}. In \cite{kandasamy_asynchronous_2017}, the authors demonstrate that TS can be easily tailored for asynchronous parallel Bayesian optimization and yields favorable comparisons with more complex techniques. Meanwhile, \cite{eriksson_scalable_2021} introduces a scalable approach to constrained Bayesian optimization leveraging GP-TS.

The performance of TS depends upon the specified probabilistic model, with potential for either under-exploration or over-exploration depending on the setting \cite{russo_tutorial_2020}. In \cite{do_epsilon-greedy_2024}, Do et al. therefore propose a scheme based on a combination of GP-TS with a $\epsilon$-greedy policy to achieve higher exploitation rates. Lastly, TS is often combined with approximate sampling schemes \cite{russo_tutorial_2020}, however, these approximations can affect the convergence behavior. The next subsection will recall techniques for approximate sampling from a GP. 

\subsection{Sampling approximate paths from Gaussian processes}\label{subsec:decoupled_sampling}
Applications relying on random realizations of GPs, such as TS, are limited by the inefficiency of generating exact draws. More specifically, two issues arise: Firstly, sampling paths from GPs scale cubically in the number of test locations \cite{wilson_efficiently_2020}. Secondly, the realizations are vectors and not functions. Thus, the test locations must be known in advance, and no gradient information is available.

In \cite{wilson_efficiently_2020}, Wilson et al. propose an approximate sampling scheme for GP posteriors to address these issues. Their path-wise GP update reads 
\begin{equation}
    \label{eq:approximate_GP_update}
    (\hat{f}_n | \boldsymbol{y})(\cdot) \overset{d}{\approx} \sum_{i=1}^l w_i\phi_i(\cdot) + \sum_{j=1}^n v_j k(\cdot,\boldsymbol{x}_j),
\end{equation}
where $\phi_i = \sigma \sqrt{\frac{2}{l}} cos(\boldsymbol{\theta}_i^T \boldsymbol{x} + \tau_i)$ denote random Fourier features (RFF), $ \boldsymbol{\theta}$ are drawn from the spectral density of the respective kernel, $\tau_i \sim \mathcal{U}(0, 2 \pi)$ and $ w_i \sim \mathcal{N}(0, 1) $ are weights that control the stochasticity of the approximate draw. The key idea of \eqref{eq:approximate_GP_update} is to combine an approximate draw of a GP prior using the RFF approximation with an exact update $\boldsymbol{v} = (\boldsymbol{K} + \sigma_\epsilon^2 \boldsymbol{I})^{-1} (\boldsymbol{y} - \hat{f}(\boldsymbol{x}) -  \boldsymbol{\epsilon}) $, with noise variates $ \boldsymbol{\epsilon} \sim \mathcal{N}(0, \sigma_{\epsilon}^2 \boldsymbol{I}) $.

The approach combines the advantages of the RFF approximation for drawing sample paths with the benefits of exact GP regression: Due to the RFF approximation of the prior, the computational cost scales linearly with the number of test locations, and the approximate draws are continuous functions. The exact update ensures the accuracy of the draws. We will follow the authors' naming convention and refer to this sampling scheme as \textit{decoupled sampling} throughout this work. 

Among the application examples considered in \cite{wilson_efficiently_2020}, GP-TS has already been considered by Wilson et al. Following their research, \cite{do_epsilon-greedy_2024} and \cite{folch_snake_2023} propose methodologies for GP-TS using decoupled sampling. A theoretical and empirical analysis of scalable GP-TS using sparse GPs and decoupled sampling is provided by \cite{vakili_scalable_2021}.

\section{Thomson Sampling for Multidisciplinary Design Optimization with Partitioned Surrogates}\label{sec:mdo_partitioned}

This section introduces the central concept of this article, an adaptation of TS to unconstrained MDO problems, called MDO-TS. 
After introducing the method, \cref{subsec:toy_problem} illustrates MDO-TS with a toy problem and \cref{sec:results} deals with the application of MDO-TS to an analytical test case of higher complexity and a numerical example.

MDO problems take the form
\begin{equation}\label{eq:MDO}
    \boldsymbol{z}^{*} = \arg \min_{\boldsymbol{z} \in Z} f_{\text{obj}}(\boldsymbol{z},\boldsymbol{y}^{*}(\boldsymbol{z}))
\end{equation}
and depend on MDA, represented as
\begin{equation}\label{eq:MDA}
    \boldsymbol{y}_i = \boldsymbol{f}_i(\boldsymbol{z}, \boldsymbol{y}_{(i)}), \, i=1,...,n_d \quad \forall \boldsymbol{z} \in Z \text{.}
\end{equation}
In \cref{eq:MDO}, $ f_{\text{obj}} $ represents the multidisciplinary objective function, while $\boldsymbol{z}$ denotes the design variables within the design space $Z$. Consequently, $ \boldsymbol{z}^* $ denotes the solution to the MDO problem. 

The converged coupling variables $\boldsymbol{y}^{*}(\boldsymbol{z})$ are determined through the solution of the MDA outlined in \cref{eq:MDA}. Within this context, $\boldsymbol{f}_i$ represents the $i$-th disciplinary model. Moreover, $\boldsymbol{y}_i$ stands for the coupling variable acquired by assessing $\boldsymbol{f}_i$, $\boldsymbol{y}_{(i)}$ symbolizes the coupling variables upon which the solution depends and $ n_d $ denotes the number of disciplines.

MDA characterized by feedback loops is referred to as a strongly coupled problem. Here, we consider strongly coupled problems and assume the existence of a unique solution for each $\boldsymbol{z} \in Z$.

\begin{figure}[htb!]
\begin{subfigure}{0.45\textwidth}
\centering
\includegraphics{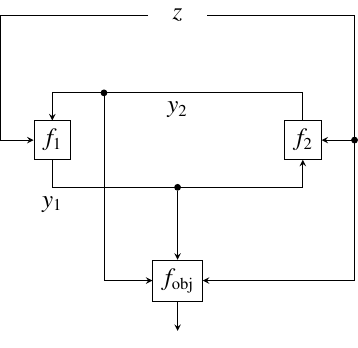}
    
    
\caption{}
\end{subfigure}
\begin{subfigure}{0.45\textwidth}
\centering
\includegraphics{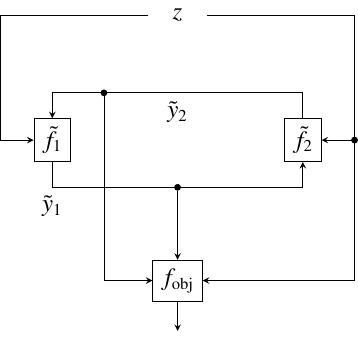}
    
    
\caption{}
\end{subfigure}
\caption{(a) MDO problem with two disciplines and feedback coupling. (b) Surrogate MDO problem where both disciplinary models are replaced by Gaussian processes.}
\label{fig:block_diagram_mdo_problem}
\end{figure}

Analogous to Dubreuil et al. \cite{dubreuil_towards_2020}, our goal is to decouple the MDA by building GP surrogates for each discipline individually, here referred to as monodisciplinary surrogates and denoted as $ \tilde{\boldsymbol{f}}_i $. \Cref{fig:block_diagram_mdo_problem} visualizes a generic MDO problem with two disciplines and feedback coupling, alongside a surrogate MDO problem where both disciplines are replaced by surrogates.
Initial surrogates are built independently of each other by an initial DoE. These surrogates are then coupled and refined by adaptively calling both solvers. For the adaptive refinement, a specific infill criterion is developed. The method is referred to as MDO-TS, since it combines coupled surrogates and TS using the RFF approximation of GPs. 

To the best of our knowledge, the sampling-based EI outlined in \cite{dubreuil_towards_2020} stands as the sole alternative infill criterion for Bayesian multidisciplinary optimization, wherein each discipline is replaced by a surrogate. Acquisition functions in Bayesian optimization rely on the uncertainty quantified by GP. By replacing each disciplinary function with a GP, uncertainty can be propagated onto the objective function through sampling. Due to the problem's nonlinearity, however, the resulting objective becomes a non-Gaussian random field. 

This principle underlies the sampling-based EI criterion proposed by Dubreuil et al. However, despite substituting disciplines with surrogates, this approach becomes computationally expensive. Each evaluation of the sampling-based EI at a selected design point $z$ requires the construction of a polynomial chaos expansion and thus a Monte Carlo simulation. In \cite{dubreuil_towards_2020}, the authors propose to set the sample size to $100$ and then to find the coefficients of the polynomial chaos expansion by ordinary least squares. A Karhunen-Loève decomposition is then performed on the resulting random field.

To migrate the resulting computational demand, the authors propose to interpolate the mean and the eigenvectors of the Karhunen-Loève expansion by GPs. This approach reduces the number of required evaluations of the sampling-based EI criterion. However, the additional interpolation introduces additional uncertainty and complexity.

There are three arguments supporting the proposed method: First, MDO-TS is less expensive to evaluate, since it utilizes the RFF approximation. Second, it provides a trade-off between exploration and exploitation for both design and coupling variables instead of solely the design variables. Third, EI is conventionally tailored to exact evaluations of the objective function. However, when each discipline is approximated by a surrogate, all we have at each step are estimates of the values of the objective. The basic idea of TS, on the other hand, is more flexible.

We propose generating and optimizing random draws, denoted as \( \hat{f}_i \), of the disciplinary surrogates \( \tilde{f}_i \). For each design point \( z \), a single MDA needs to be conducted on the random draws of the disciplinary surrogates. For the sake of completeness, it should be mentioned that the resolution of an MDA for evaluating MDO-TS is more expensive due to the required calculation of the prior path using the RFF approximation. The added expense depends linearly on the number of basis functions. However, it is decoupled from the cubic scaling with the number of observations.

Besides the obvious advantage of saving computational effort by using the RFF approximation, there is another reason why we do not use exact sample paths: Exact samples of GPs cannot replace disciplinary models in an MDA because they are vectors. Instead, we need functions that can be evaluated iteratively. One possible approach is to interpolate over the vectors. However, this method would not only be expensive, but also prone to interpolation errors. Therefore, we decided to use the decoupled sampling scheme proposed by \cite{wilson_efficiently_2020} and introduced in \cref{subsec:decoupled_sampling}.

In the following, we term the MDO problem in which disciplinary functions $ \boldsymbol{f}_i $ are substituted by GP surrogates $ \tilde{\boldsymbol{f}}_i $ as the "surrogate MDO problem." Conversely, if disciplinary functions $ \boldsymbol{f}_i $ are replaced by approximate trajectories of the surrogates $ \hat{\boldsymbol{f}}_i $, we refer to it as the "random MDO problem."

Beyond the domain $ Y $ defined by the bounds for the coupling variables, initial surrogates rely on extrapolation. While not inherently problematic, this presents two practical issues: 
Firstly, if suggested coupling variables significantly exceed the bounds, solvers may crash, impeding iterative refinement as solutions cannot be incorporated into the surrogates. Consequently, subsequent steps may encounter similar crashes.
Secondly, for field quantities that are coupling variables, dimensionality reduction is employed. The accuracy of the basis far beyond the bounds cannot be guaranteed. Therefore, evaluating the disciplinary models in these regions would result in a waste of resources. 

Hence, it is crucial to ensure that the sought solution lies within the bounded domain when designing the initial DoE. Consequently, we recommend constraining the coupling variables within their bounds during the optimization of the random or surrogate MDO problems.

The surrogate MDO problem thus becomes
\begin{equation}\label{eq:MDO_surrogate}
    \tilde{\boldsymbol{z}}^{*} = \arg \min_{\substack{\boldsymbol{z} \in Z\\\ \tilde{\boldsymbol{y}}^{*}\in Y}} f_{\text{obj}}(\boldsymbol{z},\tilde{\boldsymbol{y}}^{*}(\boldsymbol{z})) \text{,}
\end{equation}
with
\begin{equation}\label{eq:MDA_surrogate}
    \tilde{\boldsymbol{y}}_i = \tilde{\boldsymbol{f}}_i(\boldsymbol{z}, \tilde{\boldsymbol{y}}_{(i)}), \, i=1,...,n_d \quad \forall \boldsymbol{z} \in Z \text{.}
\end{equation}
The random MDO problem is defined analogously. However, the notations $ \hat{\boldsymbol{z}}^{*} $, $ \hat{\boldsymbol{y}}^{*} $, $ \hat{\boldsymbol{y}}_i $ and $ \hat{\boldsymbol{y}}_{(i)} $ are used instead of $ \tilde{\boldsymbol{z}}^{*} $, $ \tilde{\boldsymbol{y}}^{*} $, $ \tilde{\boldsymbol{y}}_i $ and $ \tilde{\boldsymbol{y}}_{(i)} $.

We suggest iterating over optimizing the random MDO problem and refining one of the disciplinary surrogates. Consequently, the disciplinary models are evaluated at the proposed design and coupling variables $\hat{\boldsymbol{z}}^{*}$ and $\hat{\boldsymbol{y}}^{*}$. Therefore, our approach offers an additional benefit beyond reducing the computational cost: It enables automatic exploration-exploitation on both the design variables and the coupling variables, whereas in \cite{dubreuil_towards_2020}, the coupling variables are chosen purely greedily.

After $n_{\text{iter}}$ iterations of the algorithm, the surrogate MDO is terminated, and the optimized design variables $\tilde{\boldsymbol{z}}^{*}$ are returned. \Cref{alg:MDO-TS} formalizes the required steps.

In \cite{cardoso_constrained_2024}, a gradient based solver combined with a multi-start strategy is used to optimize the infill criterion. Note that the decoupled sampling scheme in principle also allows for the derivation of gradients, so the same strategy would be applicable in our framework. However, we have found that the use of a genetic algorithm (GA) is sufficient to solve both the surrogate and random MDO problems due to its inexpensive evaluation for the problems we considered. Restricting the coupling variables within the bounds is accomplished by introducing a penalty term.

\begin{algorithm}
\caption{MDO using Thompson Sampling (MDO-TS)}\label{alg:MDO-TS}
\KwIn{MDO problem $ \{\boldsymbol{f}_i, \boldsymbol{f}_{obj}\} $, initial surrogates $\tilde{\boldsymbol{f}}_i$, number of iterations $ n_{\text{iter}} $, domains $ \{ Z, Y \} $}
\For{$n = 1$ to $n_{\text{iter}}$}{
    \For{$m = 1$ to $n_{d}$}{
        Draw approximate sample paths $ \hat{\boldsymbol{f}}_{1,...,n_{d}} $ from the surrogates $\tilde{\boldsymbol{f}}_{1,...,n_{d}}$\;
        Solve the random MDO problem: $ \hat{\boldsymbol{z}}^{*} = \arg \min_{\substack{\boldsymbol{z} \in Z\\\hat{\boldsymbol{y}}^{*}\in Y}} \boldsymbol{f}_{\text{obj}}(\boldsymbol{z},\hat{\boldsymbol{y}}^{*}(\boldsymbol{z})) $\;
        Store the result of the corresponding MDA: $\hat{\boldsymbol{y}}_i^* = \hat{\boldsymbol{f}}_i(\hat{\boldsymbol{z}}^{*}, \hat{\boldsymbol{y}}_{(i)}^*), \, i=1,...,n_d$ \;
        Evaluate the disciplinary function: $ \boldsymbol{y}_m = \boldsymbol{f}_m(\hat{\boldsymbol{z}}^{*}, \hat{\boldsymbol{y}}^{*}_{(m)}) $\;
        Refine the surrogate: $ \tilde{\boldsymbol{f}}_m \gets \{\{\hat{\boldsymbol{z}}^{*}, \hat{\boldsymbol{y}}^{*}_{(m)}\}, \boldsymbol{y}_m \}$\;
    }
}
Solve the surrogate MDO problem: $ \tilde{\boldsymbol{z}}^{*} = \arg \min_{\substack{\boldsymbol{z} \in Z\\\boldsymbol{y} \in Y}} \boldsymbol{f}_{\text{obj}}(\boldsymbol{z},\tilde{\boldsymbol{y}}^{*}(\boldsymbol{z})) $\;
\KwOut{Optimized design variables $ \tilde{\boldsymbol{z}}^{*} $}
\end{algorithm}

\subsection{An illustrative example}\label{subsec:toy_problem}
We follow Dubreuil et al. in that we demonstrate our approach by the 1-D toy example given by
\begin{equation}
    f_{\text{obj}}(z, y_1^{*}, y_2^{*}) = \cos \left( \frac{y_1^{*} + \exp(-y_2^{*})}{\pi} \right) + \frac{z}{20}
\end{equation}
and
\begin{equation}
    \begin{cases}  y_1 = f_1(z,y_2) = z^2 - \cos \left( \frac{y_2}{2} \right) \\ y_2 = f_2(z,y_1) = z + y_1 \end{cases} \text{.}
\end{equation}

\Cref{table:toy_problem} compares the reference solution of the toy problem with the solution of the surrogate problem before and after running MDO-TS for \nIterMDOTSToyProblem{} iterations. The initial estimate is poor, resulting in a large discrepancy in the obtained objective. Due to the simplicity of the illustrative problem, \nIterMDOTSToyProblem{} iterations are sufficient for convergence to the global optimum. 

\begin{figure}[htb!]
\begin{subfigure}{0.45\textwidth}
\includegraphics{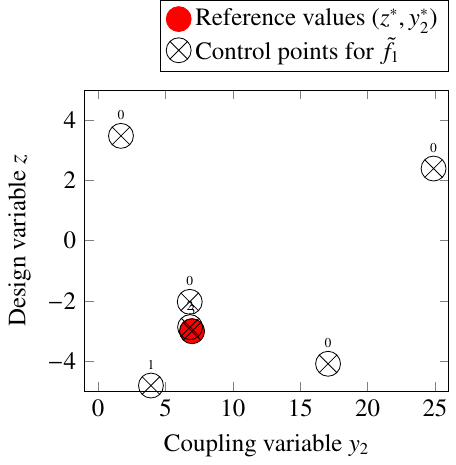}



\caption{}
\end{subfigure}
\begin{subfigure}{0.45\textwidth}
\includegraphics{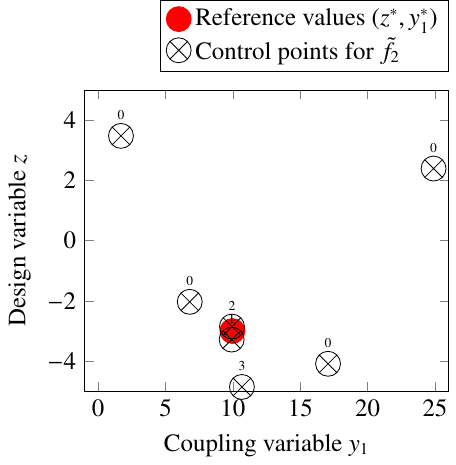}



\caption{}
\end{subfigure}

\caption{Control points for constructing the surrogates for the functions $f_1$ (a) and $f_2$ (b) of the toy problem. The reference solution is shown in red. The numbers indicate the order in which the points were added. The number "$0$" indicates that the respective point belongs to the initial DoE. The remaining points were added adaptively by MDO-TS.}

\label{fig:DoE_toy_problem}
\end{figure}

The initial Design of Experiments (DoE) consists of \sizeInitDoEToyProblem{} points, resulting in a total of $7$ evaluations of both $f_1$ and $f_2$. \Cref{fig:DoE_toy_problem} displays the initial DoE and the points added adaptively by MDO-TS. 

\begin{table}[H]
\begin{tabularx}{\textwidth}{c *6{>{\Centering}X}}
\toprule
& $z^*$ & $y_1^*$ & $y_2^*$ & $ f_{\text{obj}}(z^{*}, y_1^{*}, y_2^{*}) $ \\
\midrule
reference & \zReferenceToyProblem & \yReferenceToyProblem & \fObjReferenceToyProblem \\
Initial DoE & \zPredInitToyProblem & \yPredInitToyProblem & \fObjPredInitToyProblem \\
MDO-TS & \zPredFinalToyProblem & \yPredFinalToyProblem & \fObjPredFinalToyProblem \\
\bottomrule
\end{tabularx}
\caption{The reference solution for the toy problem, the initial estimate of the optimum, and the estimate of the optimum after \nIterMDOTSToyProblem{} iterations of MDO-TS.}
\label{table:toy_problem}
\end{table}

\definecolor{greenblue}{RGB}{0,128,128}

\begin{figure}[htb!]
\begin{subfigure}{0.45\textwidth}
\includegraphics{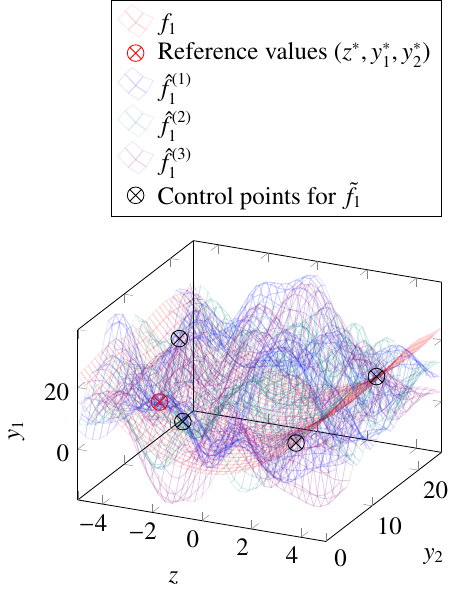}
\caption{}
\end{subfigure}
\begin{subfigure}{0.45\textwidth}
\includegraphics{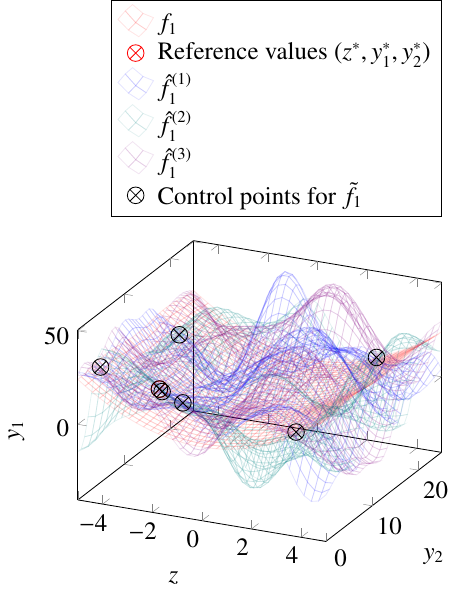}
\caption{}
\end{subfigure}
\caption{Function $f_1$ of the toy problem and three approximate random draws $ \hat{f}_1^{(1)} $, $ \hat{f}_1^{(2)} $ and $ \hat{f}_1^{(3)} $ generated with decoupled sampling before (a) and after (b) \nIterMDOTSToyProblem{} iterations of MDO-TS.}
\label{fig:draws_of_f_1}
\end{figure}

\section{Results}\label{sec:results}
In the following section, we apply MDO-TS to more challenging problems: An analytical example with three design variables and a fluid-structure interaction (FSI) example. However, the coupling variables are scalar in each case. 

To facilitate the reproducibility of the results obtained, we first provide an overview of the implementation details. To iteratively solve the surrogate MDA as well as the random MDA for specific parameters, we used the Gauss-Seidel method and accelerated the convergence with Aitken's delta-squared method. For the solution of the surrogate MDO problem or the random MDO problem, we used the differential evolution algorithm. As mentioned in \cref{sec:mdo_partitioned}, we have constrained the coupling variables by adding a penalty term. We also introduced an additional penalty term to account for cases, where the surrogate or the random coupled problems did not converge.  This occurred with increasing frequency in the later iterations of MDO-TS.  One can speculate that this is due to the increase in ill-conditioning. 

Since the decoupled sampling algorithm requires external access to the Cholesky decomposition of the covariance matrix and the kernel function, we decided to use a custom implementation of a GP with a squared-exponential kernel. We utilized the L-BFGS-B algorithm for optimizing the hyper-parameters. Thereby, the initial values for both the length-scales and the scaling of the kernel variance were set to \initScaling{}. To avoid ill-conditioning, we used a nugget for regularization and set its value to \nugget. The inputs where normalized and the outputs standardized. For both problems, we used \nBasisFunctions{} basis functions for the RFF approximation of the prior sample paths.

\subsection{Mathematical example}\label{subsec:math_example}
Analogous to \cite{dubreuil_towards_2020}, we consider an unconstrained modification of a classical MDO benchmark, the Sellar problem \cite{sellar_response_1996} with a local and a global minimum.

The modified Sellar problem is given by
\begin{equation}
    f_{\text{obj}}(\boldsymbol{z}, \boldsymbol{y}^{*}) = z_1 + z_3^2 + y_1^* + \exp(-y_2^*) + 10 \cos(z_2)
\end{equation}
and
\begin{equation}
    \begin{cases}  y_1 = f_1(\boldsymbol{z},y_2) = z_1 + z_2^2 + z_3 - 0.2y_2 \\ y_2 = f_2(\boldsymbol{z},y_1) = \sqrt{y_1} + z_1 + z_2 \end{cases} \text{.}
\end{equation}

Hereby, the design space is restricted to $ Z = [0, 10] \times [-10, 10] \times [0, 10]$ and the space of the coupling variables to $ Y = [1, 50] \times [-5, 24] $.

The size of the initial DoE was set to \sizeInitDoESellar{} and the number of iterations of MDO-TS was set to \nIterMDOTSSellar{}. Following \cite{dubreuil_towards_2020}, we used Latin hypercube sampling to generate \nRunsSellar{} initial DoEs. MDO-TS converged to the global optimum \nConvergedSellar{} times. \Cref{table:unconstrained_sellar} summarizes the results. Note that in \cite{dubreuil_towards_2020} a specific criterion was used to determine whether the functions should be evaluated at one iteration of EGMDO, while we evaluate the functions at each iteration of MDO-TS. Thus, the comparability of both approaches is limited. Dubreuil et al. ran EGMDO for $15$ iterations, but the average number of evaluations added was $8$. Setting the number of iterations much higher than $10$ without such a criterion leads to problems of ill-conditioning due to repeated, closely spaced evaluations. This is indeed an important point that we could improve. 

Another important setting is the selected tolerance of the Gauss-Seidel solver. For the computation of the reference solution, the maximum relative error was set to $10^{-10}$. However, the maximum permissible relative error for the resolution of the surrogate and the random MDA was selected more coarsely and set to $10^{-2}$. This larger relative error was allowed because we observed that at a finer resolution, the surrogates and the random MDA either converged only after a high number of iterations or not at all during the later iterations of MDO-TS. A possible explanation could be that the conditioning of the GPs deteriorates due to the repeated evaluations at nearby locations. A run is considered converged if the difference between the reference objective at $\boldsymbol{z}^*$ and the objective at $\tilde{\boldsymbol{z}}^*$ is less than $1\%$: 
\begin{equation}
    \left| \frac{f_{\text{obj}}(\boldsymbol{z}^*, \boldsymbol{y}^*) - f_{\text{obj}} \left(\tilde{\boldsymbol{z}}^*, \boldsymbol{y} \left(\tilde{\boldsymbol{z}}^* \right) \right)}{f_{\text{obj}}(\boldsymbol{z}^*, \boldsymbol{y}^*)} \right| < 0.01 
\end{equation}

Although our definition of convergence slightly differs from \cite{dubreuil_towards_2020}, the criterion presented therein would yield the same results. We observed a higher convergence rate than EGMDO as documented in \cite{dubreuil_towards_2020}, since our algorithm converged \nConvergedSellar{} instead of $88$ times. The average accuracy is also higher: While the value of the goal function is on average $0.71 \%$ off after running EGMOD, the average deviation is only $0.0458 \%$ after running MDO-TS. However, it is unclear how much of the observed differences are due to differences in basic concepts, and how much is due to differences in implementation and the difficulty of comparing the number of iterations. Furthermore, unlike Dubreuil et al. we restrict the space of the coupling variables.

\begin{table}[H]
\begin{tabularx}{\textwidth}{c *6{>{\Centering}X}}
\toprule
& $z_1$ & $z_2$ & $z_3$ & $ f_{\textrm{obj}}(\boldsymbol{z}) $ \\
\midrule
$ x $ & \zReferenceSellar & \fObjReferenceSellar \\
$ \mathbb{E}_{n=\text{value}}[x^*] $ & \zPredSellar & \fObjPredSellar \\
$ \mathbb{E}_{n=\text{value}}\left[ \lvert 100 \cdot \frac{x - x^{*}}{x} \rvert \right] $ & \zErrorSellar & \fObjErrorSellar \\
\bottomrule
\end{tabularx}
\caption{Results from \nConvergedSellar{} converged runs out of \nRunsSellar{}. The rows correspond to different values of \(x\): the first row for \(x = z_1\), the second for \(x = z_2\), the third for \(x = z_3\), and the fourth for \(x = f_{\textrm{obj}}(\boldsymbol{z})\).}
\label{table:unconstrained_sellar}
\end{table}

\subsection{2D Airfoil}\label{subsec:airfoil}
The second example consists of an engineering test case, an airfoil attached to a linear torsion spring. Hereby, we will optimize the parameterized shape of the airfoil to maximize the lift. 

\begin{figure}[htb!]
\centering
\includegraphics[width=0.5\textwidth]{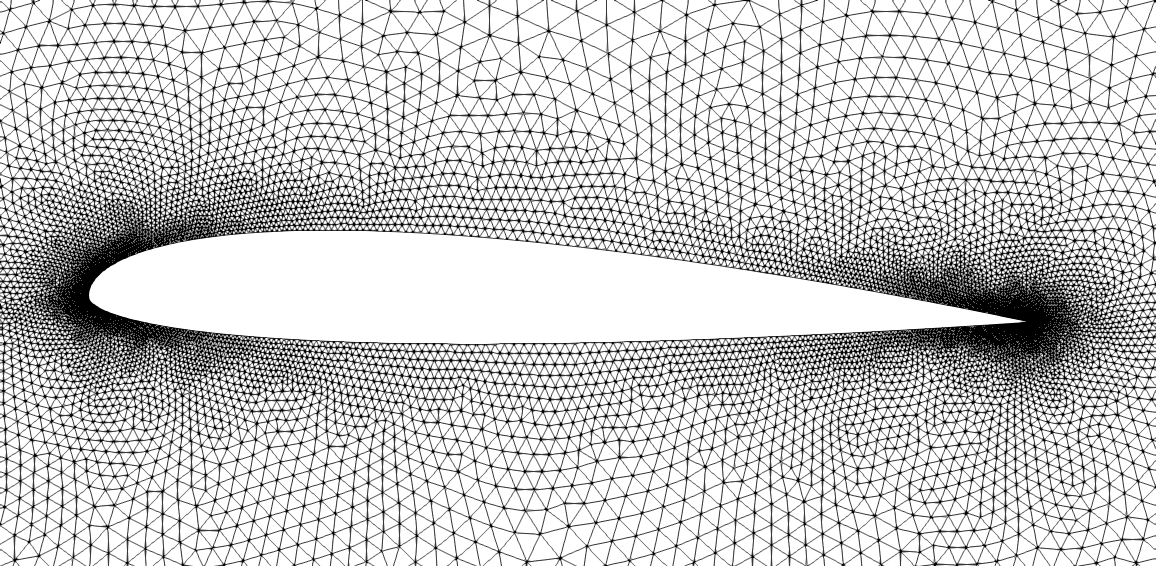}
\caption{The deformed mesh of the optimized airfoil.}
\label{fig:airfoil_mesh}
\end{figure}

The aerodynamic model is set up using the CompressiblePotentialFlowApplication from Kratos Multiphysics \cite{vicente_mataix_ferrandiz_kratosmultiphysicskratos_2024}. Specifically, we use the AlePotentialFlowSolver because it automatically calls the mesh motion solver to deform the mesh to account for both the change in angle of attack and shape changes.
The CompressiblePotentialFlowApplication is used to solve the full-potential equation,
describing a steady-state flow. Full-potential flow theory describes the velocity field as
the gradient of the velocity potential. As a result, a potential flow around an object is steady,
irrotational, and isentropic. \Cref{fig:airfoil_mesh} shows the deformed mesh of the airfoil with an optimized shape.

\begin{figure}[htb!]
\centering
\includegraphics{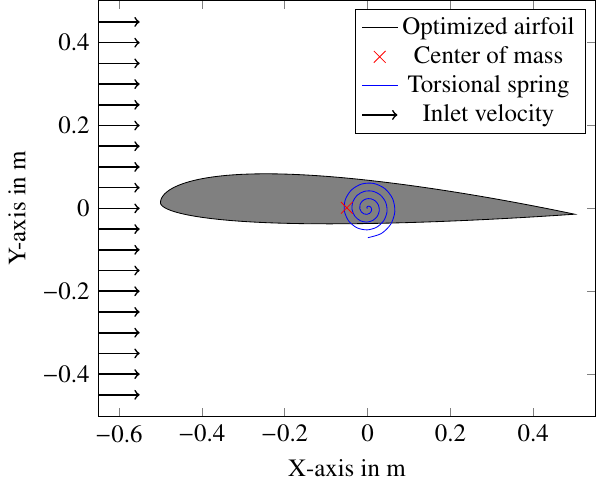}

        
        
    
\caption{The FSI problem under consideration.}
\label{fig:airfoil_setup}
\end{figure}

The initial angle of attack is set to \initAngleOfAttack, the free-stream Mach number to \machInfinity{} and the spring constant to \springConstant{}. The spring is attached to the airfoil at the origin \springCoordinates{} and the initial position of the center of mass is \centerOfMass. Since the leading edge of the original airfoil is located at $x = - 0.5$, the origin is therefore located at $0.5 \cdot c$ and the center of mass is located at $0.45 \cdot c$, where $c$ represents the chord length. The center of mass is assumed to be independent of the airfoil shape and thus only changes with the angle of attack. \Cref{fig:airfoil_setup} visualizes the setup of the experiment.

The airfoil is parameterized using Class-Shape Transformations (CST) \cite{kulfan_universal_2008} and the number of design variables for the top and bottom sides of the airfoil is set to \nCSTSide, resulting in a total number of design variables of \nCSTTotal. This relatively small number of CST coefficients was chosen to avoid overly complicating the optimization problem with a high-dimensional design space. However, since the CST parameterization is specifically designed for airfoils, we still have reasonable accuracy: The root mean square error between the original NACA 0012 profile and the fitted parametrized airfoil is only $10^{-4}$.

Since the original shape is obtained by parameterizing a symmetric NACA 0012 profile, the initial angle of attack is \initAngleOfAttack, no lift is generated initially, except for numerical errors. The CST parameters are allowed to vary within \CSTRange{} around the initial CST parameter. A range of \CSTRange{} was chosen because numerical problems were observed with a higher range. These problems were not investigated further. A possible cause could be the associated distortion of the small elements near the edges.

Thus, the design parameters are the CST variables and the coupling variables are the angle of twist and the torque. It should be noted, that the fluid surrogate needs to predict both the lift and the torque. Consequently, we build two single-output Gaussian processes and sample both outputs simultaneously. Therefore, the performance could potentially be improved by using multi-output Gaussian processes. 

Thus, the problem can be formally described by
\begin{equation}
    f_{\text{obj}}(L) = - L
\end{equation}
and
\begin{equation}
    \begin{cases}
    \phi = f_{\text{struct}}(\tau) = - \frac{\tau}{C} \\
    \begin{bmatrix} \tau \\ L \end{bmatrix} = f_{\text{aero}}(\phi, v_\text{CST}) = \delta \cdot \left( L(p(v_\text{CST})) \cdot \cos{\phi} + D(p(v_\text{CST})) \sin{\phi} \right) \end{cases}\text{.}
\end{equation}
Here, $\phi$ denotes the angle of twist, $\tau$ the torque and $C$ the spring constant. $\delta$ is the distance between the origin, where the spring is attached and the center of mass. 
$L$ and $D$ denote the lift and the drag, respectively. The notation $L(p)$ and $D(p)$ is selected to indicate that both are functions of the numerically computed pressure field $p$. The pressure field is in turn a function of the shape parameters denoted by $v_\text{CST}$. In the notation introduced in \cref{eq:MDO} and \cref{eq:MDA}, $v_\text{CST}$ corresponds to $z$, $\phi$ represents $y_1$, and $\begin{bmatrix} \tau \\ L \end{bmatrix}$ corresponds to $y_2$.

\begin{figure}[htb!]
\centering
\includegraphics{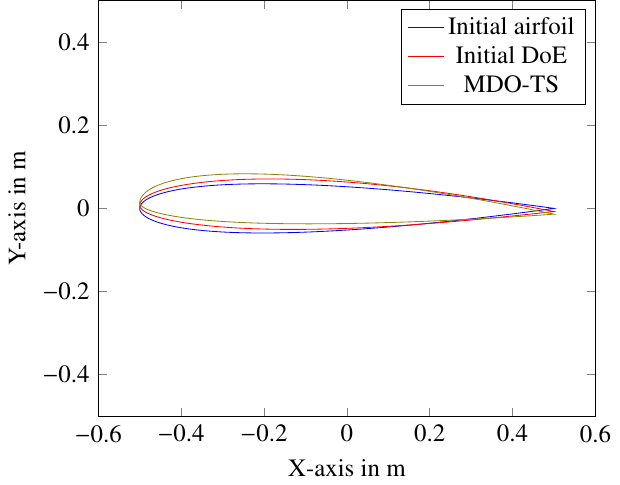}
\caption{The initial airfoil shape obtained by fitting a NACA 0012 profile, the predicted optimal shape using the initial DoE, and the optimization result after running MDO-TS.}
\label{fig:airfoil_shapes}
\end{figure}

As in case of the mathematical example, the initial size of the DoE was set to \sizeInitDoEAirfoil{} and then \nIterMDOTSAirfoil{} evaluations were added iteratively. \Cref{table:airfoil} compares the angle of attack, the torque and the lift of the original airfoil shape, the prediction of the optimal airfoil shape with the initial DoE, and the prediction of the optimal airfoil shape after running MDO-TS for \nIterMDOTSAirfoil{} iteration. Note that the optimal CST coefficients are at the upper bound of the design space as this corresponds to a wide top and a small bottom of the airfoil. Although no reference solution was computed, it seems very plausible that this solution is the global optimum with respect to the given design space. As listed in \cref{table:airfoil}, the lift could be increased significantly. \Cref{fig:airfoil_shapes} shows the obtained airfoil shapes inclined by the respective angle of attack resulting from the solution of the FSI problem.

The maximum relative error for the resolution of the surrogate and the random MDA was set again to $10^{-2}$. To obtain \cref{table:airfoil}, the original MDA was solved with a tolerance set to $10^{-6}$.

\begin{table}[H]
\begin{tabularx}{\textwidth}{c *6{>{\Centering}X}}
\toprule
& CST variables & Angle in rad & Torque in Nm & Lift in N \\
\midrule
Initial airfoil & $\begin{bmatrix}\CSTInitAirfoil\end{bmatrix}$ & \angleOfTwistInitAirfoil & \torqueInitAirfoil & \liftInitAirfoil \\
Initial DoE & $\begin{bmatrix}\CSTInitDoE\end{bmatrix}$ & \angleOfTwistInitDoE & \torqueInitDoE & \liftInitDoE \\
MDO-TS & $\begin{bmatrix}\CSTMDOTS\end{bmatrix}$ & \angleOfTwistMDOTS & \torqueMDOTS & \liftMDOTS \\
\bottomrule
\end{tabularx}
\caption{The CST variables, the values of the coupling variables (angle of attack and torque) and the value of the objective (lift) for the initial CST variables, the predicted optimal CST variables using the initial DoE, and the predicted optimal CST variables after \nIterMDOTSAirfoil{} iterations of MDO-TS.}
\label{table:airfoil}
\end{table}

\section{Discussion}\label{sec:discussion}
This paper has shown how TS can be adapted for MDO when each discipline is approximated by a GP. Compared to EGMDO, the infill criterion introduced by Dubreuil et al. \cite{dubreuil_towards_2020}, the computational overhead in the evaluation of the criterion is reduced and MDO-TS provides exploration with regard to the coupling variables. In addition, MDO-TS compared favorably with EGMDO when applied to the analytical example. However, as discussed in \cref{subsec:math_example}, caution should be exercised in drawing conclusions, as the results may not be directly comparable. EGMDO clearly outperformed other optimizers on the analytical example in terms of the number of evaluations required, as documented in \cite{dubreuil_towards_2020}. Therefore, the same can be said for MDO-TS.

Even though MDO-TS could be effectively applied to the modified Sellar problem and shape optimization of an airfoil on a spring, there are several objections to be raised against the approximation of strongly coupled systems by mono-disciplinary data-driven surrogates, both in general and specifically for optimization. 

The extension of such approaches to physical examples with coupling variables that are field quantities is an open problem. Since the coupling variables serve as inputs to the disciplinary models, and thus to the surrogates, a lower dimensional representation is necessary to mitigate the curse of dimensionality. 

Approaches to obtain such a lower dimensional representation of the coupling variables either by dimensionality reduction or by parameterization have been introduced in the literature. To obtain the representation by dimension reduction, initial evaluations are required. An obvious way to generate initial evaluations would be to evaluate the entire multidisciplinary system until convergence at an initial DoE. However, if the entire multidisciplinary system had to be solved for each design point, it is unlikely that the partitioned approach would yield any benefits: The converged coupling variables are exclusively a function of the design variables. Thus, adding the coupling variables as inputs into the surrogates provides redundant information. Consequently, the dimensionality of the input space would be artificially enlarged.

Several approaches suggest evaluating the system iteratively, but not until convergence. However, there is no reason to assume that a sample of unconverged coupling variables is sufficiently similar to a sample of converged coupling variables. An alternative would be to parameterize the coupling variables in advance. However, it is often difficult to parameterize the design space in a meaningful way. It may also be undesirable to invest in the parameterization of the coupling variables. In both cases, the additional parameterization or dimensionality reduction required introduces errors that are not present when the entire multidisciplinary system is approximated by a single data-driven surrogate.

Furthermore, the surrogate MDO problem must also be solved iteratively. Thus, the surrogates must predict outside the converged space at least well enough for an iterative solver to converge. We also found that it became more difficult to achieve convergence as more samples were added. One possible explanation could be progressive ill-conditioning of the surrogates. While Gaussian processes with smooth kernels are generally susceptible to ill-conditioning, the problem is amplified in this setting: To avoid wasting a large part of the computational budget on exploring the space of coupling variables, it is necessary to set more sample points adaptively and thus potentially close to each other.

In addition to these general drawbacks of partitioned surrogates for approximating strongly coupled multidisciplinary systems, there are also drawbacks that are more specific to optimization. These drawbacks are related to the fact that the surrogates are not built over pairs of design variables and corresponding objective evaluations. Consequently, it is more difficult to evaluate the result of the optimization compared to traditional Bayesian optimization with a single surrogate. Furthermore, crashes of disciplinary solvers cannot be penalized, although they are likely to occur when optimizing complex models.

Finally, both conceptual and implementation complexity are increased if each discipline is replaced by a surrogate compared to classical Bayesian optimization, where the entire system is replaced by a single surrogate. This means an increase in the number of settings a user has to specify, as well as an increase in potential points of failure. 

Future research should attempt to systematically address the general objections to partitioned surrogates for strongly coupled multidisciplinary systems discussed above as well as extend the scope of frameworks for multidisciplinary optimization with partitioned surrogates such as MDO-TS further. For example by adapting MDO-TS to constrained optimization analogous to \cite{cardoso_constrained_2024}. Additionally, multi-output Gaussian processes \cite{liu_remarks_2018} could be utilized to account for the correlation of the outputs of disciplinary surrogates, such as the fluid surrogate discussed in \cref{subsec:airfoil}.

\section{Conclusions}
This paper introduces "Multidisciplinary Design Optimization using Thompson Sampling", short MDO-TS, a strategy for Multidisciplinary Design Optimization (MDO) where a Gaussian process (GP) surrogate is built for each discipline of a multidisciplinary system. We have demonstrated the effectiveness of the proposed method, called MDO-TS, for two examples with scalar coupling variables, an analytical example and a numerical example. To the best of our knowledge, MDO-TS is only the second infill criterion for MDO with coupled surrogates that has been introduced in the literature. However, as discussed in \cref{sec:discussion}, there are numerous obstacles that need to be overcome before the proposed approach can be robustly scaled to problems of higher complexity, especially with non-scalar coupling variables.






\section*{Use of AI tools declaration}
The authors declare they have not used Artificial Intelligence (AI) tools in the creation of this article.

\section*{Acknowledgments}
The authors would like to acknowledge the funding by the Deutsche Forschungsgemeinschaft (DFG, German Research Foundation) under Germany’s Excellence Strategy–EXC 2163/1- Sustainable and Energy Efficient Aviation – Project ID 390881007.

\section*{Conflict of interest}
The authors have no financial or proprietary interests in any material discussed in this article.

\printbibliography

\end{document}